\documentclass[a4paper,12pt]{article}

\usepackage{latexsym}
\usepackage{bm,amsfonts,amsmath,amssymb} 
\usepackage[utf8]{inputenc}
\usepackage{abstract}
\usepackage[T1]{fontenc}
\usepackage{graphicx}
\usepackage{verbatim}
\usepackage{ulem}
\usepackage{setspace}

\usepackage{epsfig}
\usepackage{color}

\usepackage[perpage]{footmisc}
\usepackage{fancybox}
\usepackage{dsfont}

\begin{document}

\newcommand{\R}{\mathbb{R}}
\newcommand{\C}{\mathbb{C}}
\newcommand{\N}{\mathbb{N}}
\newcommand{\Z}{\mathbb{Z}}
\newcommand{\Q}{\mathbb{Q}}
\newcommand{\mj}{\mathcal}

\newcommand{\rg}{\mathrm{rg}}
\newcommand{\su}{\mathrm{supp}}
\newcommand{\la}{\lambda}
\newcommand{\fcv}{\rightharpoonup}
\newcommand{\bg}{\begin}
\newcommand{\ds}{\displaystyle}
\newcommand{\Om}{\Omega}
\newcommand{\eps}{\epsilon}
\newcommand{\Sp}{\mathbb{S}}
\newcommand{\inj}{\hookrightarrow}
\newcommand{\n}{\textbf{n}}
\newcommand{\bv}{\textbf{b}}
\newcommand{\h}{\textbf{h}}
\newcommand{\A}{\textbf{A}}
\newcommand{\F}{\textbf{F}}
\newcommand{\B}{\textbf{B}}
\newcommand{\dr}{\partial}
\newcommand{\vv}{\textbf{v}}
\newcommand{\uu}{\textbf{u}}
\newcommand{\tr}{\mathrm{Tr}}
\newcommand{\conj}{\overline}
\newcommand{\intn}{\int_{\Om}\!\!\!\!\!\!\!-}
\newcommand{\dive}{\mathrm{div}}

\newtheorem{lem}{Lemma}[section]
\newtheorem{theo}[lem]{Theorem}
\newtheorem{prop}[lem]{Proposition}
\newtheorem{sch}[lem]{Scholie}
\newtheorem{cor}[lem]{Corollary}
\newtheorem{conjec}[lem]{Conjecture}

\def\got#1{{\bm{\mathfrak{#1}}}}

\newenvironment{preuve}
{\noindent{\textbf{Proof.}}\\\rm\noindent}
{\bg{flushright}\tiny $\blacksquare$\end{flushright}}

\newenvironment{rem}
{\noindent\addtocounter{lem}{1}
{\textbf{Remark \thelem.}}\\\noindent\rm}
{\bg{flushright}\tiny $\blacksquare$\end{flushright}}

\renewcommand{\theequation}{\thesection.\arabic{equation}}
\renewcommand{\thefigure}{\thesection.\arabic{equation}}

\title{Contribution to the asymptotic analysis of the Landau-de Gennes functional}
\author{Nicolas Raymond,}
\date{}
\maketitle{}

\noindent \textit{Universit\'e Paris-Sud 11\newline 
Bâtiment 425, Laboratoire de Math\'ematiques\newline 
91405 Orsay Cedex\newline 
e-mail : nicolas.raymond@math.u-psud.fr}

\bg{abstract}
In this paper we are interested in the Landau-de Gennes functional introduced to study the transition between the smectic and nematic phases of a liquid crystal. We define a reduced functional by constraining the director field to satisfy a non-homogeneous Dirichlet condition and we prove that, below a critical temperature and if some elastic coefficients are explicitly large, then, the minimizers have to be nematic phases.
\end{abstract}

\section{Introduction and main results}
\subsubsection*{Landau-de Gennes functional}
Let $\Om$ be a connected open bounded subset of $\R^3$ with smooth boundary which represents the domain occupied by the liquid crystal. The energy of the crystal depends on a complex valued function $\psi$, called order parameter, and on the $\Sp^2$-valued vector field of the molecules denoted by $\n$ ; this energy is given by the Landau-de Gennes functional :
\bg{align*}
\mj{F}_{DG}^0(\psi,\n)&=\int_\Om|(i\nabla+q\n)\psi|^2 dx-\kappa^2\int_\Om |\psi|^2 dx+\frac{\kappa^2}{2}\int_\Om |\psi|^4 dx\\
&+K_1\int_\Om (\nabla\cdot\n)^2 dx+K_2\int_\Om|\n\cdot\nabla\times\n+\tau|^2 dx+K_3\int_\Om|\n\times(\nabla\times\n)|^2 dx\\
&+(K_2+K_4)\int_\Om\tr((\nabla\n)^2-(\nabla\cdot\n)^2) dx.
\end{align*}
\paragraph{Physical interpretation of the parameters (see \cite{Degennes})}~\\
$\kappa^2$ can be interpreted as the temperature\footnote{Actually, $\kappa^2$ is the opposite of the temperature ; in \cite{BCLP}, $\kappa^2$ was denoted by $-r$} ; $\tau$ is called the chirality (due to the default of symmetry of the molecules) and $\frac{2\pi}{q}$ corresponds to the distance between the layers of the organized phase of the liquid crystal. The $K_i$'s are called the elastic coefficients and correspond to the elastic deformations of the crystal.
\paragraph{Framework}~\\
We first assume, as in \cite{Pan,HelPan}, that 
$$K_2=K_3 \mbox{ and } K_2+K_4=0.$$
We denote $$\mj{V}(\Om)=H^1(\Om,\C)\times V(\Om,\Sp^2),$$
where $$V(\Om,\Sp^2)=\{\n\in L^2(\Om,\Sp^2): \nabla\times\n\in L^2,\,\,\nabla\cdot\n\in L^2\}.$$
We will call \textbf{phases} the elements of $\mj{V}(\Om)$.
A phase $(\psi,\n)$ such that $\psi=0$ is called a \textbf{nematic phase} and a phase such that $\psi\neq 0$ is called a \textbf{smectic phase}.\\
For positive $K_1, K_2,q,\tau,\kappa$ and $(\psi,\n)\in \mj{V}(\Om)$, we define (see \cite{Degennes,BCLP,Pan}):
\bg{align}\label{functional}
\mj{F}(\psi,\n)&=\int_{\Om}|(i\nabla+q\n)\psi|^2 dx-\kappa^2\int_\Om |\psi|^2 dx+\frac{\kappa^2}{2}\int_\Om |\psi|^4 dx\\
\nonumber&+K_1\int_\Om (\nabla\cdot\n)^2 dx+K_2\int_\Om|\nabla\times\n+\tau\n|^2dx.
\end{align}
The energy $\mj{E}(K_1,K_2,q,\tau,\kappa)$ of the functional $\mj{F}$ is defined by :
\bg{equation}\label{energy}
\mj{E}(K_1,K_2,q,\tau,\kappa)=\inf_{(\psi,\n)\in \mj{V}(\Om)}\mj{F}(\psi,\n).
\end{equation}
In order to obtain properties of the functional $\mj{F}$, Helffer and Pan have studied in \cite{HelPan} the reduced functional 
$$\mj{G}=\mj{F}_{|W(\Om)} \mbox{ with } W(\Om)=H^1(\Om,\C)\times\mj{C}(\tau),$$
and with :
$$\mj{C}(\tau)=\{\n\in L^2(\Om,\Sp^2) : \nabla\times\n+\tau\n=0\}.$$
A description of $\mj{C}(\tau)$ is provided in Appendix \ref{Ctau}.
We denote by $\mj{N}_\tau$ the set :
\bg{equation}\label{Ntau}
\mj{N}_\tau=\{(0,\n), \n\in\mj{C}(\tau)\}.
\end{equation}
We notice that the set of all nematic minimizers of $\mj{F}$ is a subset of $\mj{N}_\tau.$
Letting
\bg{equation}\label{eng}
g(q,\tau,\kappa)=\inf_{(\psi,\n)\in W(\Om)}\mj{G}(\psi,\n),
\end{equation}
the following lemma is obvious :
\bg{lem}\label{trivial}
$$\mj{E}(K_1,K_2,q,\tau,\kappa)\leq g(q,\tau,\kappa)(\leq 0).$$
\end{lem}
Moreover, the inverse inequality is asymptotically true (see \cite{HelPan}) :
\bg{equation}\label{cvg}
\lim_{K_1,K_2\to+\infty}\mj{E}(K_1,K_2,q,\tau,\kappa)=g(q,\tau,\kappa).
\end{equation}
Here, two kinds of questions appear. 
Firstly, we can wonder how $g(q,\tau,\kappa)$ depends on $(q,\tau,\kappa)$ and a partial answer is given by the following important proposition the proof of which can be found in \cite{HelPan} :
\bg{prop}\label{rtransition}
We have the equivalence :
$$\mu^*(q,\tau)\geq\kappa^2 \Leftrightarrow \left((\psi,\n) \mbox{ minimizer of $\mj{G}$} \Rightarrow \psi=0\right) \Leftrightarrow g(q,\tau,\kappa)=0,$$
with 
\bg{equation}\label{mustar}
\mu^*(q,\tau)=\inf_{\n\in\mj{C}(\tau)}\mu(q\n),
\end{equation}
where $\mu(q\n)$ denotes the lowest eigenvalue of the Neumann realization of $(i\nabla+q\n)^2$ on $\Om$.
\end{prop}
It will appear that the so-called \textbf{phase transition} is the regime of parameters such that $\kappa^2=\mu^*(q,\tau)$ and $\mu^*(q,\tau)$ is called \textbf{critical temperature}. The question of the dependence of $\mu^*(q,\tau)$ on $(q,\tau)$ will be addressed in the present work and so will be the issue of the behaviour of $g(q,\tau,\kappa)$ near the phase transition (see (\ref{ing}) in Proposition \ref{NLagmon} ).
Secondly, the authors do not obtain a control of the rate of convergence in (\ref{cvg}) ; moreover, in view of (\ref{rtransition}), it is natural to wonder if the minimizers of $\mj{F}$ are nematic when $\kappa^2\leq\mu^*(q,\tau)$ and when $K_1$ and $K_2$ are large enough. At least, we have the following obvious result (see Lemma \ref{trivial} and Proposition \ref{rtransition}) :
\bg{lem}\label{trivial2}
For all $K_1,K_2,q,\tau,\kappa>0$, if $\kappa^2>\mu^*(q,\tau)$, then all the minimizers of $\mj{F}$ are smectic.
\end{lem}
Moreover, this problem of the nematicity is related to the convergence of minimizers of $\mj{F}$. Indeed, it has been proved that, in some weak sense (see \cite{BCLP, HelPan2}), when $K_1$ and $K_2$ tend to infinity, the minimizers $(\psi,\n)$ of $\mj{F}$ tend to $(\psi^\infty,\n^\infty)$ with $\n^\infty\in\mj{C}(\tau)$ and it turns out that the crucial point is the control of this convergence. Nevertheless, putting aside the idea of an explicit control, we still have a result stating that, below the critical temperature, the minimizers are nematic for large $K_2$ :
\bg{theo}\label{trivmin0}
For all $\kappa>0$, $\tau>0$, $K_1^0>0$, there exists $\Pi(\kappa,\tau,K_1^0)>0$ such that, for all $q>0$, $K_1>K_1^0$, $K_2\geq \Pi(\kappa,\tau,K_1^0)$, if $\mu^*(q,\tau)>\kappa^2$, then the set of minimizers of $\mj{F}$ is $\mj{N}_\tau$.  
\end{theo}
\bg{rem}
Let us compare this result with those which were presented in \cite{BCLP}. 
\bg{enumerate}
\item If we put the result of Theorem \ref{trivmin0} together with the one of Lemma \ref{trivial2}, we observe that the equation $\kappa^2=\mu^*(q,\tau)$ permits
to know if the liquid crystal is smectic or nematic in the case of large $K_2$. In \cite{BCLP}, the approach is different ; in that paper, the authors compare $\kappa^2$ with some functions of the product $q\tau$ (as suggested by P-G de Gennes). More precisely, they prove that, if $\kappa^2\geq\conj{r}(q\tau)$, the minimizers are nematic phases and if $\kappa^2\leq\underline{r}(q\tau)$, the minimizers are smectic phases (see \cite[Fig. 2]{BCLP}). Moreover, they show that for small $q\tau$, these functions behave as $(q\tau)^2$ and for $q\tau$ large as $q\tau$. Actually, these behaviours are exactly the one of $\mu^*(q,\tau)$ (see \cite{Pan, HelPan, Ray} and Theorem \ref{estimatepan2}).
\item In \cite[Theorem 2]{BCLP}, the authors assume that $q\gg\tau$, which can be easily reformulated as $\sqrt{q\tau}\gg\tau$ ; this assumption is slightly weaker than in Theorem \ref{estimatepan2}, but our conclusions are stronger. So, we can say that the properties of the functions $\conj{r}$ and $\underline{r}$ just reflect the properties of the spectral quantity $\mu^*(q,\tau).$
\item We do not emphasize on the conditions on the elastic coefficients and do not try to be optimal, our interest being only in the definition of a critical temperature (and the link between $q$, $\tau$ and $\kappa$).
\end{enumerate}
\end{rem}
The result of Theorem \ref{trivmin0} is only qualitative. Thus, we are led to write the Euler-Lagrange equations satisfied by the minimizers of $\mj{F}$, hoping for some (explicit) elliptic control of the solutions. We will observe that such a control is not clear because of a rather bad Lagragian multiplier. Facing this problem in \cite[Section 5]{Pan} and \cite{Pan4}, Pan has studied, in the case when $\tau=0$, a functional with a fixed, non-homogeneous Dirichlet condition on the director field. In our work, let us first recall that $\tau$ is positive ; then, we will assume that the director field is an element of $\mj{C}_{\dr\Om}(\tau)$ on the boundary, where
$$\mj{C}_{\dr\Om}(\tau)=\{\n\in\mj{C}^{\infty}(\dr\Om) : \exists\tilde{\n}\in\mj{C}(\tau) : \n=\tilde{\n}_{|\dr\Om}\},$$ but it will not be fixed.\\ 
There are mainly two reasons to do this. The first is technical and appears when trying to get an explicit control of the convergence of minimizers when $K_1$ and $K_2$ tend to infinity. The second is "physical" ; indeed, we have recalled that the director field of the minimizers tends to some element of $\mj{C}(\tau)$. So, this is the same idea as the one which led to the reduced functional $\mj{G}$. Let us now introduce some notation.
We denote $$\mj{V}^\tau(\Om)=H^1(\Om,\C)\times V^{\tau}(\Om,\Sp^2),$$
where $$V^{\tau}(\Om,\Sp^2)=\{\n\in V^\tau(\Om,\R^3) : |\n(x)|=1\,\,a.e\},$$
and $$V^\tau(\Om,\R^3)=\{\A\in L^2(\Om,\R^3): \nabla\times\A\in L^2,\,\,\nabla\cdot\A\in L^2\,\mathrm{and}\, \A_{|\dr\Om}\in\mj{C}_{\dr\Om}(\tau)\}.$$
Our concern will be the functional $\mj{F}^{Dir}=\mj{F}_{|\mj{V}^\tau(\Om)}$ whose energy is :
$$\mj{E}^{Dir}(K_1,K_2,q,\tau,\kappa)=\inf_{(\psi,\n)\in \mj{V}^\tau(\Om)}\mj{F}^{Dir}(\psi,\n).$$
We can state our two main theorems. The first one consists of an energy estimate.
\bg{theo}\label{asymen}
For all positive $q,\tau,\kappa$ such that $\tau^2\notin\sigma(-\Delta^D)$, there exists $c_1(q,\tau,\kappa)>0$ and $c_2(q,\tau,\kappa)>0$ s.t for all $K_1, K_2$, we have :
$$g(q,\tau,\kappa)-\frac{c_1(q,\tau,\kappa)}{\sqrt{K}}-\frac{c_2(q,\tau,\kappa)}{K}\leq\mj{E}^{Dir}(K_1,K_2,q,\tau,\kappa)\leq g(q,\tau,\kappa),$$
where 
$$K=\min(K_1,K_2).$$  
\end{theo}
A choice of $c_1(q,\tau,\kappa)$ and $c_2(q,\tau,\kappa)$ will be explicited in (\ref{c1c2}).\\
In the following, we let 
\bg{equation}\label{tildeg}
\tilde{g}(q,\tau,\kappa)=g(q,\tau,\kappa)+\frac{\kappa^2|\Om|}{2}.
\end{equation}
Our second theorem states a sufficient condition to have nematic minimizers below the critical temperature.
\bg{theo}\label{trivmin}
There exists $C(\Om)>0$, such that, for all positive $q,\tau,\kappa$ satisfying $\kappa^2<\mu^*(q,\tau)$, $\tau^2\notin\sigma(-\Delta^D)$, there exists $c_3(q,\tau,\kappa)>0$ such that for all $K_1,K_2>0$, if $$K\geq C(\Om)\frac{c_3(q,\tau,\kappa)}{\sqrt{\mu^*(q,\tau)}-\kappa},$$
then the set of minimizers of $\mj{F}^{Dir}$ is $\mj{N}_\tau$. Moreover, we can take as $c_3$ :
\bg{equation}\label{c3}
c_3(q,\tau,\kappa)=q(1+\kappa)^{1/2}\tilde{g}(q,\tau,\kappa)^{1/2}\left(1+\frac{\tau^2}{\mu_\tau^1}\right)^{1/2},
\end{equation}
where $\mu_\tau^1$ defined in Section 5.1.
\end{theo}
Whereas Theorem \ref{trivmin0} was not explicit about the explosion of elastic coefficients at the phase transition, the last theorem gives an explicit lower bound of $K_1$ and $K_2$. Indeed, we observe that the right side explodes when $\kappa^2$ is close to $\mu^*(q,\tau)$ ; more precisely, for fixed $q$ and $\tau$, it is clear that $g(q,\tau,\kappa)$ tends to $0$ as $\kappa^2$ tends to $\mu^*(q,\tau)$ (see for instance Lemma \ref{lemma} and \ref{psi1}) and so $\tilde{g}(q,\tau,\kappa)$ tends to $\frac{\kappa^2|\Om|}{2}$. This fact is consistent with the physical observations (see \cite{Degennes})~; this kind of behaviour did not appear in \cite{BCLP}. 
\paragraph{Organization of the paper}~\\
This paper is organized as follows. In Section 2, we prove some properties linked with the reduced functional $\mj{G}$. In Section 3, we give the proof of Theorem \ref{trivmin0}. In Section 4, we study the Euler-Lagrange equations satisfied by the minimizers of $\mj{F}$. Finally, in Section 5, we prove Theorems \ref{asymen} and \ref{trivmin}.
\section{Properties related to the reduced functional~$\mj{G}$}
This section is devoted to the analysis of the function $(q,\tau)\mapsto\mu^*(q,\tau)$ defined in (\ref{mustar}) and to the control of the energy $g(q,\tau,\kappa)$ introduced in (\ref{eng}) near the phase transition.
\subsection{Dependence of the critical temperature $\mu^*(q,\tau)$ on $(q,\tau)$}
\paragraph{A uniform estimate when $q\tau\to+\infty$}~\\
We recall an estimate we have obtained in \cite{Ray} which permits to relax the condition $\tau$ bounded of \cite{HelPan} :
\bg{theo}\label{estimatepan}
Let $c_0>0$ and $0\leq x<\frac{1}{2}$. There exists $C>0$ and $q_0>0$ depending only on $\Om$, $c_0$ and $x$ such that, if $(q,\tau)$ verifies $q\tau\geq q_0$ and 
\bg{equation}\label{tau}
\tau\leq c_0 (q\tau)^x,
\end{equation}
then :
\bg{equation}\label{estimatepan2}
\Theta_0-\frac{C}{(q\tau)^{1/4-x/2}}\leq\frac{\mu^*(q,\tau)}{q\tau}\leq\Theta_0+\frac{C}{(q\tau)^{1/3-2x/3}},
\end{equation}
where $\Theta_0$ denotes the bottom of the spectrum of the magnetic Neumann Laplacian on $\R^3_+$ with constant magnetic field (with strength $1$). 
\end{theo}
\paragraph{Behaviour of $\mu^*(q,\tau)$ with respect to $\tau$}~\\
In this paragraph, we want to know if we have a more global information about $\mu^*(q,\tau)$ with respect to $q\tau$ as it is suggested by the last paragraph and by de Gennes (see \cite{Degennes}).
The aim of this paragraph is to prove the following Lipschitzian control :
\bg{prop}\label{lipmu}
There exists $C(\Om)>0$ such that, we have for all positive $(\tau,\tilde{\tau})$ and $q\geq 0$ :
$$|\sqrt{\mu^*(q,\tau)}-\sqrt{\mu^*(q,\tilde{\tau})}|\leq C(\Om)|q\tau-q\tilde{\tau}|$$
\end{prop}
We start by stating a general lemma :
\bg{lem}\label{contmu}
For all $A_0, A_1\in\mj{C}^{\infty}(\conj{\Om})$, we have :
\bg{equation}\label{lips}
|\sqrt{\mu(A_0)}-\sqrt{\mu(A_1)}|\leq\|A_0-A_1\|_{\infty}.
\end{equation}
\end{lem}
\bg{preuve}
Let $\psi_0$ be a $L^2$-normalized eigenfunction associated with $\mu(A_0)$ ; by the mini-max principle, we have :
$$\mu(A_1)\leq\int_\Om|(i\nabla+A_1)\psi_0|^2 dx.$$
Then, we get :
\bg{align*}
\int_\Om|(i\nabla+A_1)\psi_0|^2 dx &=\int_\Om|(i\nabla+A_0)\psi_0|^2 dx\\
                                   &+2\Re\left(\int_\Om(i\nabla+A_0)\psi_0\cdot (A_1-A_0)\conj{\psi_0}\right)+\int_\Om |A_0-A_1|^2|\psi_0|^2 dx\\
                                   &\leq \mu(A_0)+2\|A_0-A_1\|_{\infty}\sqrt{\mu(A_0)}+\|A_0-A_1\|_{\infty}^2.
\end{align*}
So, we have :
$$\mu(A_1)\leq \mu(A_0)+2\|A_0-A_1\|_{\infty}\sqrt{\mu(A_0)}+\|A_0-A_1\|_{\infty}^2=(\sqrt{\mu(A_0)}+\|A_0-A_1\|_{\infty})^2.$$
We obtain :
$$\sqrt{\mu(A_1)}\leq \sqrt{\mu(A_0)}+\|A_0-A_1\|_{\infty}.$$
Exchanging the role of $A_0$ and $A_1$, we infer (\ref{lips}).
\end{preuve}
We now define the function $\mu$ on $SO_3\times\R_+\times\R^*_+$ by :
\bg{equation}\label{mu}
\mu(Q,q,\tau)=\mu(q\n_\tau^Q),
\end{equation}
where $\n_\tau^Q$ is defined in Appendix \ref{Ctau}.
\bg{prop}
$\mu$ is continuous on $SO_3\times\R_+\times\R^*_+$ and $\mu^*$ is continuous on $\R_+\times\R^*_+$.
\end{prop}
\bg{preuve}
$(Q,q,\tau)\mapsto q\n_\tau^Q$ is clearly continuous and thus, with Lemma \ref{contmu}, we deduce that $\mu$ is continuous.
With (\ref{mustar}) and (\ref{mu}), $\mu^*(q,\tau)$ can be rewritten as : 
$$\mu^*(q,\tau)=\inf_{Q\in SO_3} \mu(Q,q,\tau),$$ we conclude that $\mu^*$ is continuous.
\end{preuve}
We now prove Proposition \ref{lipmu}.\\
Lemma \ref{contmu} provides for all $\tau>0,$ $\tilde{\tau}>0,$ $q\geq 0$ and $Q\in SO_3$ :
$$\left|\sqrt{\mu(q\n_\tau^Q)}-\sqrt{\mu(q\n_{\tilde{\tau}}^Q)}\right|\leq q\|\n_\tau^Q-\n_{\tilde{\tau}}^Q\|_\infty.$$
We apply the Taylor formula to the r. h. s to get :
$$\left|\sqrt{\mu(q\n_\tau^Q)}-\sqrt{\mu(q\n_{\tilde{\tau}}^Q)}\right|\leq C(\Om)q|\tau-\tilde{\tau}|.$$
Thus, we can write :
$$\sqrt{\mu^*(q,\tau)}\leq\sqrt{\mu(q\n_\tau^Q)}\leq \sqrt{\mu(q\n_{\tilde{\tau}}^Q)}+C(\Om)q|\tau-\tilde{\tau}|.$$
Taking the infimum over $Q$ of the right term, we deduce :
$$\sqrt{\mu^*(q,\tau)}\leq\sqrt{\mu^*(q,\tilde{\tau})}+C(\Om)q|\tau-\tilde{\tau}|.$$
Exchanging the roles of $\tau$ and $\tilde{\tau}$, Proposition \ref{lipmu} is proved.
\subsection{Estimate of $g(q,\tau,\kappa)$ near the phase transition}
In this subsection, we give an energy estimate for the reduced functional $\mj{G}$ when $q\tau$ is large and near the phase transition $\kappa^2=\mu^*(q,\tau)$, in the smectic domain. Let us emphasize that we will not assume that $\tau$ stays in a bounded interval $]0,\tau_0[$.
Let us now state a lemma (for a proof, cf. \cite{HelPan}) :
\bg{lem}\label{lemma}
Let $(\psi,\n)$ be a minimizer of $\mj{G}$, then :
\bg{equation}\label{g}
g(q,\tau,\kappa)=-\frac{\kappa^2}{2}\int_{\Om}|\psi|^4 dx,
\end{equation}
and :
\bg{equation}\label{controlpsi2}
\int_\Om |\psi|^4\leq \left(1-\frac{\mu(q\n)}{\kappa^2}\right)\int_{\Om}|\psi|^2 dx.
\end{equation}
\end{lem}
Let us observe that, if $\kappa^2<\mu^*(q,\tau)$, the r.h.s of (\ref{controlpsi2}) is zero and so, the minimizers are zero.\\
Moreover, the following lemma is proved in \cite{DGP} and \cite[Section 11.3]{FouHel2} :
\bg{lem}\label{psi1}
For all minimizers $(\psi,\n)$ of $\mj{G}$, we have :
\bg{equation}
\|\psi\|_{\infty}\leq 1.
\end{equation}
\end{lem}
The next proposition permits to estimate $\|\psi\|_4$ and to see the exponential decrease of $\psi$ away from the boundary in the case where $\tau$ is not necessarily bounded.
\bg{prop}\label{NLagmon}
For $x\in[0,\frac{1}{2}[$, $c_0>0$ and $b\in]\Theta_0,1[$, there exists $\sigma_0>0$, $C>0$ and $\alpha>0$ such that for all $(q,\tau,\kappa)$ s.t $q\tau\geq \sigma_0$, $\tau\leq c_0(q\tau)^x$ and ${\frac{\kappa^2}{b}\leq\mu^*(q,\tau)<\kappa^2}$, we have for all $(\psi,\n)$ minimizer of $\mj{G}$ :
$$\int_{\Om} \left\{e^{\alpha\sqrt{q\tau}t(x)}|\psi|^2+\frac{1}{q\tau}|(i\nabla+q\n)\psi|^2\right\} dx\leq\frac{C}{\sqrt{q\tau}}$$
and
\bg{equation}\label{controlpsi}
\int_\Om|\psi|^4 dx\leq \frac{C}{\sqrt{q\tau}},
\end{equation}
where we have let :
$$t(x)=d(x,\dr\Om).$$
Moreover, under the same hypotheses, we have the estimate :
\bg{equation}\label{ing}
-C\frac{(\kappa^2-\mu^*(q,\tau))^2}{\kappa^2\sqrt{q\tau}}\leq g(q,\tau,\kappa)\leq 0.
\end{equation}
\end{prop}
\bg{preuve}
Denoting by $\mu_0^{\Om}(q,\A)$ the lowest eigenvalue of the Dirichlet realization of $(i\nabla+q\A)^2$ on $\Om$, with the uniform estimate of $\mu_0^{\Om}(q\tau,\frac{\n}{\tau})$ with $\n\in\mj{C}(\tau)$ obtained in \cite{Ray} and implementing the non linear Agmon estimates (cf. \cite{Agmon, HelMo2}), we get~:
\bg{equation}\label{agmon}
\int_{\Om} \left\{e^{\alpha\sqrt{q\tau}t(x)}|\psi|^2+\frac{1}{q\tau}|(i\nabla+q\n)\psi|^2\right\} dx\leq C\int_\Om |\psi|^2 dx.
\end{equation}
Let us recall the proof to see precisely the dependence on $\tau$.
We use the identity :
\bg{align*}
\|\left(i\nabla+q\n\right)e^{\alpha\sqrt{q\tau}t}\psi\|^2_2-\alpha^2 q\tau\| |\nabla t| e^{\alpha\sqrt{q\tau}t}\psi\|_2^2&=\kappa^2\|e^{\alpha\sqrt{q\tau}t}\psi\|_2^2-\kappa^2\|e^{\alpha\sqrt{q\tau}t}|\psi|^2\|_2^2\\
&\leq \kappa^2\|e^{\alpha\sqrt{q\tau}t}\psi\|_2^2.
\end{align*}
We let $u=e^{\alpha\sqrt{q\tau}t}\psi$ and introduce for a given $r>0$ a partition of unity (associated to a covering by balls of centers $x_j$ and radius $r$) as in \cite{HelMo} :
\bg{align*}
&\sum_j \chi_j^2=1\\
&\sum_j |\nabla\chi_j|^2\leq\frac{C}{r^2}.
\end{align*}
Then, the IMS formula gives (see \cite{Cycon}) :
$$\|\nabla_{q\n}u\|^2_2\geq\sum_j \|\nabla_{q\n}\chi_j u\|^2_2-\frac{C}{r^2}\|u\|^2_2,$$
where we have let : 
$$\nabla_{q\n}=i\nabla+q\n.$$
We deduce :
$$\sum_{j\mathrm{int}} \left(\|\nabla_{q\n}\chi_j u\|^2_2-(\kappa^2+\alpha^2 q\tau+\frac{C}{r^2})\|\chi_j u\|^2_2\right)\leq(\kappa^2+\alpha^2 q\tau+\frac{C}{r^2})\sum_{j\mathrm{bnd}}\|\chi_j u\|^2_2.$$
We use Corollary 5.4 of \cite{Ray} to get :
$$\|\nabla_{q\n}\chi_j u\|^2_2\geq (q\tau-C(q\tau)^{3/4+x/2})\|\chi_j u\|^2_2.$$
The crucial point which permits to relax the condition "$\tau$ bounded" (which was done in the previous literature, see for example \cite{HelPan}) is that, under the assumption $x<\frac{1}{2}$, we have $\frac{3}{4}+\frac{x}{2}<1$.
Thus, we find, letting $\ds{r=\frac{\eps_0}{\sqrt{q\tau}}}$, with $\eps_0$ and $\alpha$ small enough :
$$\sum_{j\mathrm{int}}\|\chi_j u\|^2_2\leq C(b,\alpha,\eps_0)\sum_{j\mathrm{bnd}}\|\chi_j u\|^2_2.$$
The end of the proof of (\ref{agmon}) is standard and left to the reader ; it uses Lemma \ref{psi1}.
Finally, with Lemma \ref{lemma} we find :
$$\int_\Om|\psi|^4 dx\leq\int_\Om |\psi|^2 dx\leq C\int_{t\leq 2\eps_0(q\tau)^{-1/2}}|\psi|^2 dx\leq \frac{C}{(q\tau)^{1/4}}\left(\int_\Om|\psi|^4 dx\right)^{1/2}$$
and the control of $\psi$ in $L^4$ follows.\\
Finally, we notice that :
$$1-\frac{\mu(q\n)}{\kappa^2}\leq 1-\frac{\mu^*(q,\tau)}{\kappa^2}.$$
and we have necessarily $\kappa^2>\mu(q\n)$ ; if not, the minimizers would be trivial (see Lemma \ref{lemma}).
Using (\ref{controlpsi2}), we get :
\bg{align*}
\int_\Om|\psi|^4 dx&\leq\left(1-\frac{\mu^*(q,\tau)}{\kappa^2}\right)\int_\Om|\psi^2| dx\\
                   &\leq C \left(1-\frac{\mu^*(q,\tau)}{\kappa^2}\right)\int_{t\leq 2\eps_0 (q\tau)^{-1/2}}|\psi|^2 dx\\
                   &\leq \tilde{C}\left(1-\frac{\mu^*(q,\tau)}{\kappa^2}\right)\frac{1}{(q\tau)^{1/4}}\left(\int_\Om|\psi|^4 dx\right)^{1/2}.
\end{align*}
This improves (\ref{controlpsi}) in 
$$\int_\Om |\psi|^4 dx \leq \tilde{C}^2\left(1-\frac{\mu^*(q,\tau)}{\kappa^2}\right)^2\frac{1}{(q\tau)^{1/2}}$$
and using (\ref{g}), we obtain (\ref{ing}).
\end{preuve}
\section{Minimizers of $\mj{F}$ for large $K_1$ and $K_2$ for $\kappa^2$ below the critical temperature}
This section deals with the proof of Theorem \ref{trivmin0}.
It is standard that $\mj{F}$ admits minimizers (the proof is slightly different from the one of \cite{BCLP} because in our work $K_2+K_4=0$).
Then, we can write that :
\bg{align}
\mj{F}(\psi,\n)=\int_{\Om}|(i\nabla+q\n)\psi|^2 dx-\frac{\kappa^2}{2}|\Om|+\frac{\kappa^2}{2}\int_\Om(|\psi|^2-1)^2 dx\\
\nonumber+K_1\int_\Om (\nabla\cdot\n)^2 dx+K_2\int_\Om|\nabla\times\n+\tau\n|^2 dx.
\end{align}
Using Lemma \ref{trivial}, we can get for any $(\psi,\n)$ minimizer of $\mj{F}$ the following upper bounds :
\bg{align}
\label{first}\int_\Om |\nabla\times\n+\tau\n|^2 dx\leq \frac{\tilde{g}(q,\tau,\kappa)}{K_2},\\
\label{first2}\int_\Om (\nabla\cdot\n)^2 dx\leq\frac{\tilde{g}(q,\tau,\kappa)}{K_1},\\
\label{twice}\int_\Om(|\psi|^2-1)^2 dx\leq\frac{2\tilde{g}(q,\tau,\kappa)}{\kappa^2},\\
\label{third}\int_{\Om}|(i\nabla+q\n)\psi|^2 dx\leq\tilde{g}(q,\tau,\kappa),
\end{align}
where $\tilde{g}(q,\tau,\kappa)$ is defined in (\ref{tildeg}).
Moreover, let us recall that (\ref{psi1}) still holds.

\subsection{Nematicity of the minimizers of $\mj{F}$}
We are now interested in the regime $\kappa^2\leq\mu^*(q,\tau)$ and in the proof of Theorem \ref{trivmin0}.
\paragraph{Case when $\kappa=0$}~\\
Let us briefly notice what happens when $\kappa=0$. The functional becomes :
$$\mj{F}(\psi,\n)=\int_\Om|(i\nabla+q\n)\psi|^2 dx+K_1\int_\Om (\nabla\cdot\n)^2 dx+K_2\int_\Om|\nabla\times\n+\tau\n|^2 dx.$$
Clearly, the phases $(0,\n)$, with $\n\in\mj{C}(\tau)$ are minimizers of $\mj{F}$ with energy $0$. Moreover, if $(\psi,\n)$ is a minimizer of $\mj{F}$, it implies that :
$$\nabla\cdot\n=0 \mbox{ and } \nabla\times\n+\tau\n=0$$
and thus $\n\in\mj{C}(\tau)$. In addition, we have $$\int_\Om|(i\nabla+q\n)\psi|^2 dx=0$$
which provides, by the diamagnetic inequality $\int_\Om|\nabla|\psi||^2 dx=0$ and thus $|\psi|$ is constant (equal to $c$).
If $c\neq 0$, we write $\psi=ce^{i\phi}$, we find $q\n=\nabla\phi$, thus $\nabla\times \n=0=-\tau\n$ and this is a contradiction. Gathering all these remarks, we deduce :
\bg{prop}
In the case when $\kappa=0$, the set of the minimizers of $\mj{F}$ is $\mj{N}_\tau$ (cf. (\ref{Ntau})).
\end{prop}
\paragraph{Proof of Theorem \ref{trivmin0}}~\\
What follows is inspired by \cite{BCLP}. We show in this paragraph that, if $K_1$ and $K_2$ are "large" enough, then, the minimizers are nematic phases.\\
The next proposition deals with the behaviour of the director field of the minimizers when $K_2$ is large. 
\bg{prop}\label{approxn}
For all $\eps>0$ and for all $\kappa\neq 0$, $\tau>0$, $K_1^0>0$, there exists $\Pi(\kappa,\tau,K_1^0)>0$ such that for all $K_1>K_1^0$, $K_2\geq\Pi(\kappa,\tau,K_1^0)$, $q>0$, and for all $(\psi,\n)$ minimizer of $\mj{F}$, there exists $\tilde{\n}\in\mj{C}(\tau)$ such that, we have : 
$$\|\n-\tilde{\n}\|_{L^4(\Om)}\leq\eps.$$
\end{prop}
\bg{preuve}
The proof follows from (\ref{first}), (\ref{first2}) and from \cite[Lemma 3.4]{HelPan2}.
\end{preuve}
We will also use the following lemma :
\bg{lem}\label{inq}
For any $(\psi,\n)$ minimizer of $\mj{F}$ (or $\mj{F}^{Dir}$), we have :
$$\int_{\Om} |(i\nabla+q\n)\psi|^2 dx\leq\kappa^2\int_\Om |\psi|^2 dx.$$
\end{lem}
\bg{preuve}
If $\psi=0$, this is trivial. If $\psi\neq 0$ and if the converse inequality were true, we would find that $0\geq\mj{F}(\psi,\n)>0$ and this would be a contradiction.
\end{preuve}
We can now prove Theorem \ref{trivmin0}.
Let $(\psi,\n)$ be a minimizer of $\mj{F}$.
We write, for some $\tilde{\n}$ to be choosen later :
\bg{align*}
\|(i\nabla+q\n)\psi\|_{L^2(\Om)}^2&\geq\|(i\nabla+q\tilde{\n})\psi\|_{L^2(\Om)}^2\\
&-2\|(i\nabla+q\tilde{\n})\psi\|_{L^2(\Om)}\|(\n-\tilde{\n})\psi\|_{L^2(\Om)}+q^2\|(\n-\tilde{\n})\psi\|^2_{L^2(\Om)}.
\end{align*}
Moreover, we have :
$$\|(\n-\tilde{\n})\psi\|_{L^2(\Om)}\leq\|\n-\tilde{\n}\|_{L^4(\Om)}\|\psi\|_{L^4(\Om)}$$
and :
$$\|(i\nabla+q\tilde{\n})\psi\|_{L^2(\Om)}\leq \|(i\nabla+q\n)\psi\|_{L^2(\Om)}+q\|(\n-\tilde{\n})\psi\|_{L^2(\Om)}.$$
We infer, with Lemma \ref{inq}, that :
\bg{align*}
\kappa^2\|\psi\|^2_{L^2(\Om)}&\geq\mu^*(q,\tau)\|\psi\|^2_{L^2(\Om)}\\
&-2q\kappa\|\psi\|_{L^2(\Om)}\|\psi\|_{L^4(\Om)}\|\n-\tilde{\n}\|_{L^4(\Om)}-q^2\|\n-\tilde{\n}\|^2_{L^4(\Om)}\|\psi\|^2_{L^4(\Om)}.
\end{align*}
By the Sobolev embedding, we have first : 
$$\||\psi|\|_{L^4(\Om)}\leq C(\Om)\||\psi|\|_{H^1(\Om)}.$$
Then, the diamagnetic inequality provides :
$$\|\nabla|\psi|\|_{L^2(\Om)}\leq\|(i\nabla+q\n)\psi\|_{L^2(\Om)}\leq\kappa\|\psi\|_{L^2(\Om)}.$$
Consequently, we have :
$$\|\psi\|_{L^4(\Om)}\leq C(\Om)(1+\kappa)\|\psi\|_{L^2(\Om)}.$$
By contradiction, we assume that $\psi\neq 0$ and we obtain :
$$\kappa^2-\mu^*(q,\tau)+2C(\Om)q\kappa\sqrt{1+\kappa}\|\n-\tilde{\n}\|_{L^4(\Om)}+C(\Om)^2q^2(1+\kappa)\|\n-\tilde{\n}\|^2_{L^4(\Om)}\geq 0.$$
We now look at the second degree trinomial which appears and we get that, for some $C(\Om)>0$, if
$$\|\n-\tilde{\n}\|_{L^4(\Om)}\leq C(\Om)\frac{-q\kappa\sqrt{1+\kappa}+\sqrt{q^2\kappa^2(1+\kappa)+4q^2(1+\kappa)(\mu^*(q,\tau)-\kappa^2)}}{q^2(1+\kappa)}$$
then a contradiction occurs.\\
So, choosing $\eps>0$ such that :
$$\eps<C(\Om)\frac{-q\kappa\sqrt{1+\kappa}+\sqrt{q^2\kappa^2(1+\kappa)+4q^2(1+\kappa)(\mu^*(q,\tau)-\kappa^2)}}{q^2(1+\kappa)},$$
and as $\tilde{\n}$ the field provided by Proposition \ref{approxn}, we have proved Theorem \ref{trivmin0}.

\section{Euler-Lagrange equations for the minimizers of $\mj{F}$}
The aim of this section is to show that we cannot a priori deduce from the Euler-Lagrange equations a quantitative version of Proposition \ref{approxn}, even in the simplified case when $K_1=K_2$.
More precisely, we study the Euler-Lagrange equation obtained after differentiation with respect to $\n$ satisfied by
each minimizer $(\psi,\n)$ of $\mj{F}$ (see \cite{Pan})~:
\bg{align}
\label{eli2}&\langle T,\uu\rangle=0\,\, \mathrm{for\, all}\, \uu\in H^1_0(\Om,\R^3) \,\,\mathrm{s.t}\,\, \uu\cdot\n=0,
\end{align}
where $$T=-K_1\nabla(\nabla\cdot\n)+K_2(\nabla\times+\tau)^2\n-2q\Im(\conj{\psi}\nabla\psi).$$
We now use the identity :
$$\nabla\times(\nabla\times\n)=-\Delta\n+\nabla(\nabla\cdot\n)$$
and find :
$$T=(K_2-K_1)\nabla(\nabla\cdot\n)+K_2(-\Delta\n+2\tau\nabla\times\n+\tau^2\n)-2q\Im(\conj{\psi}\nabla\psi).$$
As $\n\in H^1(\Om)$, we get $\Delta\n\in H^{-1}(\Om)$ and $\nabla(\nabla\cdot\n)\in H^{-1}(\Om)$. Moreover, as $\n\in H^1(\Om,\Sp^2)$ and $\|\psi\|_\infty\leq 1$, we notice that $T\in H^{-1}(\Om)$ and so, we can define the products of distribution $T\cdot\n$ and $T\times\n$ as follows.\\
For all $\psi\in\mj{C}_0^{\infty}(\Om,\R^3)$ and $\phi\in\mj{C}_0^{\infty}(\Om)$, we let :
$$\langle T\times\n,\psi\rangle=\langle T,\n\times\psi\rangle,$$
$$\langle T\cdot\n,\phi\rangle=\langle T,\phi\n\rangle.$$
In addition, these two distributions can be extended to continuous linear forms on $H^1_0(\Om)$.
\bg{lem}\label{lap}
In the sense of distributions, we have 
$$-\Delta\n\cdot\n=|\nabla\n|^2.$$
\end{lem}
Noticing (double exterior product formula) that, for all $S\in\mj{D}'(\Om)$, we have~:
$$S=\n\times (S\times\n)+(S\cdot\n)\n,$$
we get :
\bg{lem}\label{S}
Let $S$ be in $H^{-1}(\Om)$. 
If $S\cdot\n=0$ and $S\times\n=0$, then $S=0.$
\end{lem}
We deduce the following proposition :
\bg{prop}
If $K_1=K_2$, then $T\cdot\n\in L^1(\Om)$ and $T=(T\cdot\n)\n$.
\end{prop}
\bg{preuve}
The first statement comes from Lemma \ref{lap}, and from the property that $\nabla\times\n\in L^2(\Om)$ and $|\nabla\n|^2\in L^1(\Om)$.
Let us prove the second one.\\ 
We can now define : $\tilde{T}=(T\cdot\n)\n$. We let $S=T-\tilde{T}$. Then, $S$ satisfies the assumptions of Lemma \ref{S} and thus $S=0.$
\end{preuve}
We obtain the following corollary :
\bg{cor}
When $K_1=K_2$, there exists a function  $\la\in L^1(\Om)$ such that the equation (\ref{eli2}) is equivalent to
\bg{equation}\label{eli2'}
-\Delta\n+2\tau\nabla\times\n-2q\Im(\conj{\psi}\nabla\psi)=\la \n,\,\mathrm{in}\,\Om.
\end{equation}
In particular, $\Delta\n\in L^1(\Om).$
\end{cor}
So, this is not enough to have an elliptic control of the minimizers. Actually, we can prove that the minimizers are $\mj{C}^\infty$ almost everywhere in $\conj{\Om}$ (see \cite{HKL}).

\section{Minimizers of $\mj{F}^{Dir}$}
Before analysing the minimizers of $\mj{F}^{Dir}$, we need some spectral theory.
\subsection{Spectral theory}
We introduce an important operator $T_{\tau}$.
\paragraph{Injectivity of $T_{\tau}$}~\\
We denote by $\sigma(-\Delta^D)$ the spectrum of the Dirichlet Laplacian $-\Delta^D$ on $\Om$ and we assume that 
\bg{equation}\label{sa}
\tau^2\notin\sigma(-\Delta^D). 
\end{equation}
We introduce the quadratic form on $H^1_0(\Om,\R^3)$ :
$$Q_{\tau}(u)=\|\nabla\cdot u\|^2_{L^2(\Om)}+\|\nabla\times u+\tau u\|^2_{L^2(\Om)}.$$
The associated operator, with domain $H^2(\Om,\R^2)\cap H^1_0(\Om,\R^3)$, is denoted by $T_{\tau}$ which can be expressed as :
\bg{equation}\label{Ttau}
T_{\tau}=-\Delta+2\tau\nabla\times+\tau^2,
\end{equation}
We denote its lowest eigenvalue by $\mu^1_{\tau}$ and we have (using the mini-max principle) :
\bg{equation}\label{lbTtau}
Q_\tau(u)\geq \mu^1_{\tau}\|u\|^2_{L^2}\quad \forall u\in H^1_0(\Om,\R^3) 
\end{equation}
As $T_{\tau}$ is elliptic of order $2$, we have the following proposition :
\bg{prop}
If $T_{\tau} u\in L^2(\Om,\R^3)$ and $u\in H^1_0(\Om,\R^3)$, then $u\in H^2(\Om,\R^3).$
\end{prop}
So, we infer that :
$$\|T_{\tau} u\|_{L^2}\geq \mu^1_{\tau}\|u\|_{L^2}\quad \forall u\in H^1_0(\Om,\R^3)\,s.t.\,T_{\tau}u \in L^2(\Om,\R^3).$$
\bg{lem}
If (\ref{sa}) is satisfied, then $\mu^1_{\tau}>0$ and $T_{\tau}$ is injective.
\end{lem}
\bg{preuve}
If $\mu^1_{\tau}=0$, then, we would immediately infer that the corresponding eigenfunction $u$ satisfies :
\bg{equation}\label{ctau0}
\nabla\cdot u=0\mbox{ and } \nabla\times u+\tau u=0.
\end{equation}
Taking the curl of the second equation in (\ref{ctau0}) and using the first, we would find :
$$-\Delta u-\tau^2 u=0.$$
With Assumption (\ref{sa}), we would obtain $u=0$ and a contradiction.
\end{preuve}
We can now reformulate (\ref{lbTtau}) by stating the following proposition :
\bg{prop}[Control of $\|u\|_{L^2(\Om)}$]\label{uL2}~\\
For all $u\in H^1_0(\Om,\R^3)$, we have :
$$\|u\|^2_{L^2(\Om)}\leq \frac{1}{\mu_\tau^1}\left(\|\nabla\cdot u\|^2_{L^2(\Om)}+\|\nabla\times u+\tau u\|^2_{L^2(\Om)}\right).$$
\end{prop}
In fact, we have a control in $H^1(\Om)$ :
\bg{prop}[Control of $\|u\|_{H^1(\Om)}$]\label{uH1}~\\
There exists $C(\Om)>0$ such that :
$$\|u\|^2_{H^1(\Om)}\leq C(\Om)\left(1+\frac{\tau^2}{\mu_\tau^1}\right)\left(\|\nabla\cdot u\|^2_{L^2(\Om)}+\|\nabla\times u+\tau u\|^2_{L^2(\Om)}\right).$$
\end{prop}
\bg{preuve}
As a consequence of the identity $-\Delta+\nabla(\nabla\cdot)=\nabla\times\nabla\times$, we have :
$$\|\nabla u\|^2_{L^2(\Om)}=\|\nabla\cdot u\|_{L^2(\Om)}^2+\|\nabla\times u\|_{L^2(\Om)}^2.$$
Moreover, we get :
\bg{align*}
\|\nabla u\|^2_{L^2(\Om)}&=\|\nabla\times u+\tau u\|_{L^2(\Om)}^2-2\tau\langle\nabla\times u,u\rangle-\tau^2\|u\|_{L^2(\Om)}^2+\|\nabla\cdot u\|_{L^2(\Om)}^2.\\
&\leq Q_\tau(u)+2\tau\|\nabla\times u\|_{L^2(\Om)}\|u\|_{L^2(\Om)}-\tau^2\|u\|_{L^2(\Om)}^2\\
&\leq Q_\tau(u)+2\tau\|\nabla u\|_{L^2(\Om)}\|u\|_{L^2(\Om)}-\tau^2\|u\|_{L^2(\Om)}^2
\end{align*}
Thus, for all $\gamma>0$, we infer :
$$\|\nabla u\|^2_{L^2(\Om)}\leq Q_\tau(u)+\tau\left(\gamma\|\nabla u\|^2_{L^2(\Om)}+\frac{1}{\gamma}\|u\|^2_{L^2(\Om)}\right)-\tau^2\|u\|^2_{L^2(\Om)}.$$
For $\tau>0$, we let $\gamma=\frac{1}{2\tau}$ and we find :
$$\frac{1}{2}\|\nabla u\|^2_{L^2(\Om)}\leq Q_\tau(u)+\frac{\tau^2}{\mu_\tau^1}Q_\tau(u).$$
\end{preuve}

\paragraph{Lower bound for $\mu^1_{\tau}$}~\\
We now want to give an explicit lower bound for $\mu^1_{\tau}$.
Let us consider $u^1_\tau$ a $L^2$-normalized eigenfunction of $T_{\tau}$ associated with $\mu^1_{\tau}$. We get :
$$\|\nabla\cdot u_\tau^1\|_{L^2}^2+\|\nabla\times u_\tau^1+\tau u_\tau^1\|_{L^2}^2=\mu^1_{\tau}.$$
Thus, we have :
$$\|\nabla\times u_\tau^1+\tau u_\tau^1\|_{L^2}\leq \sqrt{\mu_{\tau}^1}.$$
An easy computation provides :
$$-\Delta-\tau^2=(\nabla\times+\tau)^2-\nabla(\nabla\cdot)-2\tau(\nabla\times+\tau)=T_{\tau}-2\tau(\nabla\times+\tau).$$
So, we obtain :
$$(-\Delta-\tau^2)u_{\tau}^1=\mu_{\tau}u_\tau^1-2\tau(\nabla\times u_\tau^1+\tau u_\tau^1).$$
We apply the spectral theorem to have :
$$d(\tau^2,\sigma(-\Delta^D))\leq\|(-\Delta-\tau^2)u_\tau^1\|_{L^2}.$$
It follows that :
$$\mu_{\tau}^1+2\tau\sqrt{\mu_{\tau}^1}-d(\tau^2,\sigma(-\Delta^D))\geq 0.$$
Consequently, the following proposition is proved :
\bg{prop}\label{lbmu}
We have :
$$\mu^1_{\tau}\geq -\tau+\sqrt{\tau^2+d(\tau^2,\sigma(-\Delta^D))}.$$
\end{prop}
The next proposition gives the behaviour of $\mu_\tau^1$ when $\tau$ tends to 0.
\bg{prop}\label{lbmu2}
Denoting by $\la_1^D$ the lowest eigenvalue of the Dirichlet Laplacian, we have :
$$\lim_{\tau\to 0}\mu^1_{\tau}=\la_1^D.$$
\end{prop}
\bg{preuve}
We consider $\psi_0$ a $L^2$-normalized eigenfunction associated with $\la_1^D$.
$$Q_\tau(\psi_0)=\|\nabla\psi_0\|^2_{L^2(\Om)}+2\tau\int_\Om\psi_0\cdot\nabla\times\psi_0 dx+\tau^2\leq \la_1^D+2\tau\sqrt{\la_1^D}+\tau^2.$$
Thus, we deduce : $$\mu_\tau^1\leq\la_1^D+2\tau\sqrt{\la_1^D}+\tau^2.$$
In addition, we have observed that :
$$\mu_\tau^1\geq-2\tau\sqrt{\mu_\tau^1}+d(\tau^2,\sigma(-\Delta^D)).$$
But, for $\tau<\la_1^D$, we have : $d(\tau^2,\sigma(-\Delta^D))=\la_1^D-\tau^2$ and so, the result is proved.
\end{preuve}

\subsection{Trace of the elements of $V(\Om,\Sp^2)$}
In this short part, we recall how to define the trace of an element in $L^2$ whose divergence and curl are also in $L^2$ (cf. \cite{GiRa}). Thus, $\mj{F}^{Dir}$ will be well defined. 
So we start, by a density lemma (which can be proved by cut-off and regularization) :
\bg{lem}
We let : 
$$V=\{\n\in L^2(\Om,\R^3) : \nabla\cdot\n\in L^2(\Om,\R^2), \nabla\times\n\in L^2(\Om,\R^3)\},$$
and for $\n\in V$, we define the norm : 
$$\|\n\|_V^2=\|\n\|_2^2+\|\nabla\cdot\n\|_2^2+\|\nabla\times\n\|_2^2.$$
Then, $(V,\|\cdot\|)$ is an Hilbert space in which $\mj{C}^{\infty}(\conj{\Om},\R^3)$ is dense.
\end{lem}
The following proposition permits to define the trace of an element of $V$ :
\bg{prop}
Let us consider $\n\in V.$
Then, the trace of $\n$ on $\dr\Om$ is well defined as an element of $H^{-1/2}(\dr\Om,\R^3)$.
\end{prop}
\bg{preuve}
The proof is standard (see \cite{GiRa}) but we recall it briefly for completeness.
We first assume that $\n\in\mj{C}^{\infty}(\conj{\Om})$.
Let us recall some formulas ; for all $\phi\in\mj{C}^{\infty}(\conj{\Om})$ and $\uu\in\mj{C}^{\infty}(\conj{\Om},\R^3)$ : 
\bg{align}
\label{div}&\langle \n,\nabla\phi \rangle=-\langle\nabla\cdot\n,\nabla\phi\rangle-\langle \n\cdot\nu,\phi_{|\dr\Om}\rangle,\\
\label{curl}&\langle\nabla\times\n,\uu\rangle=\langle\n,\nabla\times\uu\rangle-\langle \n\times\nu,\uu_{|\dr\Om}\rangle,
\end{align}
Then, (\ref{div}) and (\ref{curl}) imply that 
$$\phi\mapsto \langle \n\cdot\nu,\phi_{|\dr\Om}\rangle\quad\mathrm{and}\quad\uu\mapsto \langle \n\times\nu,\uu_{|\dr\Om}\rangle$$
can be extended in continuous linear forms respectively on $H^1(\Om,\C)$ and $H^1(\Om,\R^3).$
Then, as $\dr\Om$ is smooth, there exists a continous operator from $H^{1/2}(\dr\Om)$ (respectively $H^{1/2}(\dr\Om,\R^3)$) to $H^{1}(\Om)$ (respectively $H^{1}(\Om,\R^3)$) denoted by $T$ such that for all $f\in H^{1/2}(\dr\Om)$, $F=Tf$ satisfies $F_{|\dr\Om}=f$.\\
It follows from the previous lemma that when $\n\in V$, we can define : 
$$\n\cdot\nu\in H^{-1/2}(\dr\Om)\quad\mathrm{and}\quad \n\times\nu\in H^{-1/2}(\dr\Om,\R^3).$$
In the case where $\n$ is regular, with the double exterior product formula, we have on $\dr\Om$ :
$$\n=\nu\times(\n\times\nu)+(\n\cdot\nu)\nu.$$
Thus, by density, we can define the trace of $\n$ on $\dr\Om$ when $\n\in V$ as the element of $H^{-1/2}(\dr\Om,\R^3)$ :
$$\n_{|\dr\Om}=\nu\times(\n\times\nu)+(\n\cdot\nu)\nu.$$
\end{preuve}

\subsection{Existence of minimizers of $\mj{F}^{Dir}$}
Using \cite[Lemma 3.6]{GiRa}, we infer :
$$\mj{V}^\tau(\Om)\subset H^1(\Om,\C)\times H^1(\Om,\Sp^2).$$
Then, let us state a lemma concerning the minimization set~:
\bg{lem}
$\mj{V}^\tau(\Om)$ is weakly compact in $H^1(\Om,\C)\times H^1(\Om,\Sp^2).$   
\end{lem}
\bg{preuve}
This is enough to prove that $V^\tau(\Om,\Sp^2)$ is weakly compact in $H^1(\Om,\Sp^2)$. Let $(\n^j)$ a weakly convergent sequence of $V^\tau(\Om,\Sp^2)$ ; we denote by $\n^\infty$ its limit.
By compact injection, we deduce that there exists a subsequence such that $\n^j$ strongly converges to $\n^{\infty}$ in $L^2(\Om)$ and so $\n^j$ pointwise converges to $\n^{\infty}$ up to another extraction. Thus, we have $|\n^{\infty}|=1$. The trace of $\n^{\infty}$ on $\dr\Om$ is well defined as a element of $H^{-1/2}(\dr\Om)$. Moreover, we can write : $\n^j_{|\dr\Om}={\n_{\tau}^{Q^j}}_{|\dr\Om}$ for some $Q^j\in SO(3)$. Up to another subsequence extraction, we can assume that $Q^j$ tends to $Q^{\infty}\in SO(3)$. Thus, we get that $\n^j_{|\dr\Om}$ uniformly converges to ${\n_\tau^{Q^{\infty}}}_{|\dr\Om}$.
\end{preuve}
We can now infer the existence of minimizers. We oberve that $(\ref{first})$, $(\ref{first2})$, $(\ref{twice})$, $(\ref{third})$ and $(\ref{psi1})$ still hold and thus, any mimimizing sequence $(\psi^j,\n^j)$ is bounded in $H^1(\Om,\C)\times H^1(\Om,\Sp^2)$. 
With the previous lemma, after a subsequence extraction, we can assume that $(\psi^j,\n^j)$ converges to $(\psi^\infty,\n^\infty)\in\mj{V}^\tau(\Om)$ and the conclusion is standard.\\
\bg{rem}
We observe that $W(\Om)\subset\mj{V}^\tau(\Om)$ and so Lemma \ref{trivial} and the inequalities (\ref{first}), (\ref{first2}), (\ref{twice}) and (\ref{third}) are still true for $\mj{F}^{Dir}$.
\end{rem}

\subsection{Energy estimate : proof of Theorem \ref{asymen}}
Let $(\psi,\n)$ be a minimizer of $\mj{F}^{Dir}$. By our choice of the domain of the functional $\mj{F}^{Dir}$, there exists $\n^Q_\tau$ such that $\n_{|\dr\Om}={\n_\tau^Q}_{|\dr\Om}$.
Then, we get by Proposition \ref{uL2} :
$$\|\n-\n_\tau^Q\|^2_{L^2(\Om)}\leq\frac{1}{\mu^1_{\tau}}\left(\|\nabla\cdot(\n-\n_\tau^Q)\|^2_{L^2(\Om)}+\|\nabla\times(\n-\n_\tau^Q)+\tau(\n-\n_\tau^Q)\|^2_{L^2(\Om)}\right).$$
We infer that :
$$\|\n-\n_\tau^Q\|^2_{L^2(\Om)}\leq\frac{1}{\mu^1_{\tau}}\left(\|\nabla\cdot\n\|^2_{L^2(\Om)}+\|\nabla\times\n+\tau\n\|^2_{L^2(\Om)}\right).$$
By (\ref{first}) and (\ref{first2}), we have :
$$\|\n-\n_\tau^Q\|^2_{L^2(\Om)}\leq\frac{1}{\mu^1_{\tau}}\left(\frac{\tilde{g}(q,\tau,\kappa)}{K_1}+\frac{\tilde{g}(q,\tau,\kappa)}{K_2}\right)\leq \frac{2\tilde{g}(q,\tau,\kappa)}{\min(K_1,K_2)\mu^1_{\tau}}.$$
In addition, we have :
$$\mj{F}^{Dir}(\psi,\n)\geq\mj{F}(\psi,\n_\tau^Q)-2q\|(i\nabla+q\n_\tau^Q)\psi\|_{L^2(\Om)}\|(\n-\n_\tau^Q)\psi\|_{L^2(\Om)}+q^2\|(\n-\n_\tau^Q)\psi\|^2_{L^2(\Om)}.$$
Moreover, we recall that $\|\psi\|_{\infty}\leq 1$ ; so, writing that~:
$$\|(i\nabla+q\n_\tau^Q)\psi\|_{L^2(\Om)}\leq\|(i\nabla+q\n)\psi\|_{L^2(\Om)}+q\|\n-\n_\tau^Q\|_{L^2(\Om)},$$
we deduce from Lemma \ref{inq} :
$$\mj{F}^{Dir}(\psi,\n)\geq g(q,\tau,\kappa)-2q\kappa|\Om|^{1/2}\|\n-\n_\tau^Q\|_{L^2(\Om)}-q^2\|\n-\n_\tau^Q\|^2_{L^2(\Om)}$$
and Theorem \ref{asymen} is proved, with 
\bg{equation}\label{c1c2}
c_1(q,\tau,\kappa)=q\kappa\left(\frac{2\tilde{g}(q,\tau,\kappa)}{K\mu_\tau^1}\right)^{1/2}\mbox{ and }c_2(q,\tau,\kappa)=\frac{2|\Om|q^2\tilde{g}(q,\tau,\kappa)}{K\mu_\tau^1}.
\end{equation}
Combining this result with \cite[Theorem 7.5]{HelPan} and Proposition \ref{lbmu2}, we get an asymptotics when $\tau$ tends to $0$ :
\bg{prop}
For all $q_0,\kappa_0>0$ and $c_0>0$, there exists ${C(q_0,\kappa_0,c_0)>0}$ and $\tau_0>0$, if $(K_1,K_2,q,\tau,\kappa)$ satisfies
$0\leq q\leq q_0$, $0\leq\kappa\leq\kappa_0$, $0\leq\tau\leq\tau_0$ and $K_1,K_2\geq c_0$, then 
$$\left|\mj{E}^{Dir}(K_1,K_2,q,\tau,\kappa)+\frac{\kappa^2|\Om|}{2}\right|\leq C(q_0,\kappa_0,c_0)\tau.$$
\end{prop}

\subsection{Nematicity/Smecticity of minimizers}
This section deals with the proof of Theorem \ref{trivmin}.
\paragraph{Smecticity for $\kappa^2>\mu^*(q,\tau)$}~\\
As a consequence of (\ref{trivial}), we observe that, without condition on $K_1$ and $K_2$, we have :
$$\mj{E}^{Dir}(K_1,K_2,q,\tau,\kappa)<0.$$
Consequently, in this case, the minimizers $(\psi,\n)$ are smectic phases.\\
We are now interested in the converse regime.
\paragraph{Nematicity for $\kappa^2<\mu^*(q,\tau)$ : proof of Theorem \ref{trivmin}}~\\
Let $(\psi,\n)$ be a minimizer of $\mj{F}^{Dir}$.
With the Cauchy-Schwarz inequality, we get :
\bg{align*}
\int_{\Om} |(i\nabla+q\n)\psi|^2 dx&\geq \|(i\nabla+q\n_\tau^Q)\psi\|^2_{L^2(\Om)}\\
&-2q\|(i\nabla+q\n_\tau^Q)\psi\|_{L^2(\Om)}\|(\n-\n_\tau^Q)\psi\|_{L^2(\Om)}+q^2\|(\n-\n_\tau^Q)\psi\|^2_{L^2(\Om)}.
\end{align*}
With Lemma \ref{inq}, we observe that :
\bg{align*}
\|(i\nabla+q\n_\tau^Q)\psi\|_{L^2(\Om)}&\leq \|(i\nabla+q\n)\psi\|_{L^2(\Om)}+q\|(\n-\n_\tau^Q)\psi\|_{L^2(\Om)}\\
&\leq\kappa\|\psi\|_{L^2(\Om)}+q\|(\n-\n_\tau^Q)\psi\|_{L^2(\Om)}.
\end{align*}
Then, the analysis is exactly the same as in Section 3 and, if $\psi\neq 0$, this leads to :
$$\kappa^2-\mu^*(q,\tau)+2C(\Om)q\kappa\sqrt{1+\kappa}\|\n-\n_\tau^Q\|_{H^1(\Om)}+C(\Om)^2q^2(1+\kappa)\|\n-\n_\tau^Q\|^2_{H^1(\Om)}\geq 0.$$
But, we have, by Proposition \ref{uH1} :
\bg{align*}
\|\n-\n_\tau^Q\|^2_{H^1(\Om)}&\leq C(\Om)\left(1+\frac{\tau^2}{\mu_\tau^1}\right)\left(\|\nabla\cdot \n\|^2_{L^2(\Om)}+\|\nabla\times \n+\tau \n\|^2_{L^2(\Om)}\right).\\
&\leq C(\Om)\left(1+\frac{\tau^2}{\mu_\tau^1}\right)\left(\frac{2\tilde{g}(q,\tau,\kappa)}{K}\right),
\end{align*}
where $K=\min(K_1,K_2)$.
Considering the second degree trinomial which appears, we get that for some $C(\Om)>0$, if 
$$\frac{1}{K}\leq C(\Om)\frac{\sqrt{\mu^*(q,\tau)}-\kappa}{q(1+\kappa)^{1/2}\tilde{g}(q,\tau,\kappa)^{1/2}\left(1+\frac{\tau^2}{\mu_\tau^1}\right)^{1/2}},$$
then a contradiction follows and Theorem \ref{trivmin} is proved.
\paragraph{Case when $\kappa^2=\mu^*(q,\tau)$}~\\
The following lemma is a consequence of the Euler-Lagrange equations :
\bg{lem}
For all $(\psi,\n)$ minimizer of $\mj{F}^{Dir}$, we have :
$$\int_\Om|(i\nabla+q\n)\psi|^2 dx=\kappa^2\left(\int_\Om |\psi|^2 dx-\int_\Om |\psi|^4 dx\right).$$
\end{lem}
Thus, we find :
\bg{align*}
\kappa^2(\|\psi\|_{L^2}^2-\|\psi\|^4_{L^4})&\geq \|(i\nabla+q\n_\tau^Q)\psi\|^2_{L^2(\Om)}\\
&-2q\|(i\nabla+q\n_\tau^Q)\psi\|_{L^2(\Om)}\|(\n-\n_\tau^Q)\psi\|_{L^2(\Om)}+q^2\|(\n-\n_\tau^Q)\psi\|^2_{L^2(\Om)}.
\end{align*}
With the Cauchy-Schwarz inequality, we obtain :
$$-\kappa^2\|\psi\|^4_{L^4}+2q\kappa\|\psi\|_{L^2}\|\psi\|_{L^4}\|\n-\n_\tau^Q\|_{H^1(\Om)}+q^2\|\n-\n_\tau^Q\|_{H^1(\Om)}^2\|\psi\|_{L^4}^2\geq 0.$$
Noticing that $\|\psi\|_{L^2(\Om)}\leq|\Om|^{1/2}\|\psi\|_{L^4(\Om)}$ and assuming that $\psi\neq 0$, we infer :
$$\|\psi\|_{L^4}^2\leq\frac{2q\kappa|\Om|^{1/2}+q^2\|\n-\n_\tau^Q\|_{H^1(\Om)}}{\kappa^2}\|\n-\n_\tau^Q\|_{H^1(\Om)}.$$ 
Consequently, we have proved the following proposition :
\bg{prop}
There exists $C(\Om)>0$ such that for all $(q,\tau,\kappa)$ satisfying $\kappa^2=\mu^*(q,\tau)$ and (\ref{sa}), there exists $c_4(q,\tau,\kappa)>0$ s.t we have for any minimizer $(\psi,\n)$ of $\mj{F}^{Dir}$ :
$$\|\psi\|_{L^4(\Om)}\leq C(\Om)\frac{c_4(q,\tau,\kappa)}{K^{1/4}},$$
where
$$c_4(q,\tau,\kappa)=\frac{q^{1/2}\tilde{g}(q,\tau,\kappa)^{1/4}}{\kappa^{1/2}}\left(1+\frac{\tau^2}{\mu_\tau^1}\right)^{1/4}.$$
\end{prop}
\appendix
\section{Description of $\mj{C}(\tau)$}\label{Ctau}
In this section, we wish to give a new proof (coming from discussions with François Alouges) of the following proposition (see \cite[Lemma 3]{BCLP}) :
\bg{prop}
Denoting by $SO(3)$ the group of the rotations of $\R^3$ and defining, for $\tau>0$ :
$$\n_{\tau}(x_1,x_2,x_3)={}^t(\cos(\tau x_3),\sin(\tau x_3),0),$$
we have :
$$\mj{C}(\tau)=\{Q\n_\tau({}^tQ),\quad Q\in SO(3)\},$$
where, for all $x\in\Om$ : 
$$\n_\tau({}^tQ) (x)=\n_\tau({}^tQ\,x).$$
\end{prop}
So, we are interested in the equation :
\bg{equation}\label{helico}
\nabla\times\n+\tau\n=0 \mbox{ with } \n\in L^2(\Om,\Sp^2).
\end{equation}
\paragraph{Rank of the Jacobian}~\\
The following lemma is consequence of \cite{BCLP} :
\bg{lem}
Any solution $\n\in H^1(\Om,\Sp^2)$ of (\ref{helico}) (with $\tau\neq 0$) is analytic and verifies $\rg(\nabla_x\n)=1$ for all $x\in\Om$.
\end{lem}
\paragraph{Local solution}~\\
The constant rank Theorem implies that there exists $\phi$ and $\psi$ two $\mj{C}^1$-diffeomorphisms in a neighborhood of $(0,0,0)$ such that :
$$\n(\psi(X_1,X_2,X_3))=\phi(0,0,X_3).$$
Thus, after differenciation, we can write locally for $p$ and $q$ $\mj{C}^{\infty}$ functions :
$$\nabla\n=\tau p\otimes q,$$
with $|p|=1$ ; let us notice that we cannot a priori assume that $|q|=1$.
As $\n$ is in $\ker(\nabla\n)$, we get $\n\cdot p=\n\cdot q=0$. As 
$$\dive(\n)=\tau p\cdot q=0$$ 
and 
$$\tau^2=|\nabla\n|^2=\tau^2 |q|^2,$$ 
we find that $(p,q,n)$ is an orthonormal basis.\\
We recall that :
$$\dive(\nabla\n)=\Delta\n=-\tau^2\n$$ 
and we have for all $j$ :
\bg{equation}\label{divpq} 
\dive(p_j q)=p_j\dive(q)+\nabla p_j q.
\end{equation}
Thus, multiplying by $p_j$, summing and remembering that $\n\cdot p=0$ and $|p|^2=1$, we find $$\dive(q)=0.$$
We now observe that $\nabla\times(p_j q)=0$ for all $j$ and that :
\bg{equation}\label{curlpq}
\nabla\times(p_j q)=\nabla p_j\times q+p_j\nabla\times q.
\end{equation}
One multiplies by $p_j$, sums and find :
$$\nabla\times q=0.$$
We conclude that $\Delta q=0$. Taking the scalar product with $q$ and noticing that $|q|=1$, we find $|\nabla q|^2=0$ and $q$ is constant.
\paragraph{End of the proof}~\\
We are reduced to search a local solution of 
$$\nabla\times\n+\tau\n=0$$
with $\n$ orthogonal to a constant direction ; and this is easy to see that such solutions can be expressed as :
$$\n=Q\n_{\tau}({}^t Q),$$
where $Q$ denotes a rotation.
Indeed, we may assume that this orthogonal direction is $e_3$. Then we can write $\n=(n_1,n_2,0)$.
The equation (\ref{helico}) becomes :
$$\left\{
\bg{array}{ccc}
-\dr_3 n_2+\tau n_1&=&0\\
\dr_3 n_1+\tau n_2&=&0\\
\dr_1 n_2-\dr_2 n_1&=&0\\
\end{array}
\right..$$
We deduce that :
$$\dr_3^2 n_i+\tau^2n_i=0,$$
for $i\in\{1,2\}$ and we find that $\n=\n_{\tau}$.
Then, by analycity, we get :
$$\n=Q\n_{\tau}({}^t Q).$$
\paragraph{Case when $\tau=0$}~\\
In this subsection, we just want to show that the properties of $\mj{C}(0)$ are very different from the one of $\mj{C}(\tau)$ with $\tau>0$. We wish to study the equation :
$\nabla\times\n=0$ with $\n\in H^1(\Om,\Sp^2)$.
If we look at the result in the case where $\tau>0$, it would suggest a family of (constant) solutions : $\{Qe_3, Q\in SO_3\}=\Sp^2$. We prove here that this set doesn't contain all the solutions.\\
For $a\notin\conj{\Om}$, we let $\ds{\n_a(x)=\frac{x-a}{|x-a|}}$. It is clear that $\n_a\in\mj{C}^{\infty}(\conj{\Om},\Sp^2).$
Using the formula :
$$\nabla\times \alpha\uu =\nabla\alpha\times\uu+\alpha\nabla\times\uu,$$
we find :
$$\nabla\times\n_a=-\frac{x-a}{|x-a|^3}\times (x-a)+\frac{1}{|x-a|}\nabla\times (x-a)=0.$$
Consequently, $\n_a$ is an element of $\mj{C}(0)$.

\paragraph{Acknowledgments}~\\
I am very grateful to Professor B. Helffer for his advices and comments. I would also like to thank F. Alouges for useful discussions concerning the results of Appendix A.

\bibliographystyle{plain}
\bibliography{biblio}
\end{document}